\documentclass[A4,useAMS,usenatbib]{mn2e} 

\usepackage{hyperref} 
\usepackage{graphicx} 
\usepackage{times} \voffset=-0.8in

\newcommand{\tpos}{\langle\delta\phi_{\parallel}\rangle_{20}} \newcommand{\spara}{\delta\sigma_{\phi,\parallel}} \newcommand{\srad}{\delta\sigma_{R_{\rm aps}}}
\newcommand{\rapo}{R_{\rm apo}} \newcommand{\vapo}{V_{\rm apo}}

\newcommand{\seq}{\,{=}\,}

\def\kms{ {\rm km~s\textsuperscript{-1}}} \def\Lz{ L_{\rm z}} \def\Iz{ I_{\rm z}} \def\msun{ {\rm M_\odot}}

\defcitealias{petersen16a}{PWK16} \defcitealias{petersen18a}{Paper I} \defcitealias{petersen18b}{Paper II} \defcitealias{petersen18c}{Paper III} \defcitealias{petersen18e}{Paper V} \defcitealias{petersen18d}{Paper IV}

\begin{document}

\title[Torque in Barred Galaxy Models] {Using Torque to Understand Barred Galaxy Models}

\author[Petersen, Weinberg, \& Katz] {Michael~S.~Petersen,$^{1,2}$\thanks{michael.petersen@roe.ac.uk} Martin~D.~Weinberg,$^2$ Neal~Katz$^2$\\ $^1$Institute for Astronomy, University of Edinburgh, Royal Observatory, Blackford Hill, Edinburgh EH9 3HJ, UK \\ $^2$Department of Astronomy, University of Massachusetts at Amherst, 710 N. Pleasant St., Amherst, MA 01003}

 \maketitle \begin{abstract} We track the angular momentum transfer in $n$-body simulations of barred galaxies by measuring torques to understand the dynamical mechanisms responsible for the evolution of the bar-disc-dark matter halo system. We find evidence for three distinct phases of barred galaxy evolution: assembly, secular growth, and steady-state equilibrium. Using a decomposition of the disc into orbital families, we track bar mass and angular momentum through time and correlate the quantities with the phases of evolution. We follow the angular momentum transfer between particles and identify the dominant torque channels. We find that the halo model mediates the assembly and growth of the bar for a high central density halo, and the outer disc mediates the assembly and growth of the bar in a low central density halo model. Both galaxies exhibit a steady-state equilibrium phase where the bar is neither lengthening nor slowing. The steady-state equilibrium results from the balance of torque between particles that are gaining and losing angular momentum. We propose observational metrics for barred galaxies that can be used to help determine the evolutionary phase of a barred galaxy, and discuss the implications of the phases for galaxy evolution as a whole. \end{abstract}

\begin{keywords} galaxies: Galaxy: halo---galaxies: haloes---galaxies: kinematics and dynamics---galaxies: evolution---galaxies: structure \end{keywords}

\section{Introduction} \label{sec:introduction}

The non-local transfer of angular momentum in galaxies gives rise to the most spectacular large-scale feature of spiral galaxies: bars. The bar feature owes to a redistribution of angular momentum in the galaxy as it reorganises stars from the exponential disc into an energetically favoured bar. Thus, a bar is both a manifestation of past evolution as well as a harbinger of possible evolution to come. Fundamentally, theorists have built their understanding of the disc-bar-dark matter halo system evolution through the transport of angular momentum, beginning with the seminal work of \citet{lynden72}.

Bars comprise a long-term (many Gyr) secularly-evolving perturbation that drains angular momentum from the inner disc. Because the bar transports angular momentum outward as a structure that trails the circular frequency, the bar becomes stronger when it loses angular momentum, giving rise to the picture that bars grow, lengthen, and slow over the course of their evolution \citep{kalnajs71,lynden72}. The strength (or amplitude) of the bar , therefore, depends on how efficiently the bar can shed angular momentum to willing sinks, such as the outer disc or dark matter halo. Canonical wisdom \citep{tremaine84, weinberg85, hernquist92, athanassoula96, debattista98, debattista00, athanassoula02, athanassoula03, holleyb05, sellwood06a} states that bars will slow their pattern speed $\Omega_p$ by depositing angular momentum into a spherical component such as the dark matter halo. Evolution in the dark matter halo will change the rate at which angular momentum may be accepted, or even stop the transfer of angular momentum altogether \citep[][hereafter PWK16]{petersen16a}. The key properties of bars are their amplitude, length, and pattern speed, which may be determined observationally (see examples of determining the length and amplitude of bars by stellar mass in \citealt{munoz13} and see \citealt{tremaine84a} for a method to determine bar pattern speeds). However, simulations indicate that the length and pattern speed of bars do not follow a simple trend that one can associate with age or evolutionary status, but are instead a complex combination of myriad parameters, primarily pertaining to the position-velocity phase-space distribution of mass in the total system. To make progress in understanding the secular evolution resulting from bars, we require a deeper understanding of the physical mechanisms that ultimately result in the observed features of barred galaxies. Understanding the channels for angular momentum transport, and the mechanisms giving rise to these channels, is imperative to interpreting and predicting future evolution from the current states of barred galaxies (e.g. observed morphology).

Unique locations in the phase-space of a galaxy where an individual orbit may change its conserved quantities (energy, $E$, and angular momentum, $\Lz$) control the transfer of angular momentum. Those locations surround resonances, where orbits may gain or lose significant $\Lz$ over a handful of rotation periods owing to low-integer commensurabilities between orbital frequencies $\Omega$, which we understand in the three cylindrical coordinates $(r,\phi,z)$, and $\Omega_p$, the pattern speed of the global mode. The most well-known resonance is corotation (CR), where the orbital frequency $\Omega_\phi$ equals that of the pattern $\Omega_p$. More generally, resonant (or commensurate) orbits satisfy the equation \begin{equation} l_1\Omega_r + l_2\Omega_\phi+l_3\Omega_z = m\Omega_p, \label{eq:resonances} \end{equation} where $(l_1,l_2,l_3)$ is a triple of small integers (usually $l_{1,2,3}\leq3$) and $m$ is the multiplicity of the pattern. The quadrupole $m\seq2$ corresponds to a bar or two-arm spiral. CR for the bar may then be represented as $(l_1,l_2,l_3)=(2,2,0)$. Other common resonances are the inner Lindblad resonance, ILR (-1,2,0), and the outer Lindblad resonances, OLR (1,2,0).  The disc transfers angular momentum to the halo at these resonant couples, and has been studied extensively in the literature (see e.g. \citealt{lynden72, tremaine84, hernquist92, weinberg02, ceverino07, weinberg07a, weinberg07b}). To an observer sitting on the bar, resonant (or commensurate) orbits trace closed, non-axisymmetric paths. Because the orbits are non-axisymmetric, they can be torqued by the bar. Once an orbit is trapped, its angular momentum $\Lz$ tends to change with the slowing of the bar as the bar transfers angular momentum to the dark matter halo and/or outer disc. In this case, the changing bar pattern speed allows resonances to sweep through frequency space such that new orbits satisfy equation~(\ref{eq:resonances}) and possibly become trapped into libration with the bar pattern. Generally, to trap into a self-gravitating pattern, one of the quantities in the pattern must be changing, such as the angular momentum (which is directly related to the pattern speed). The bar pattern continues to slow as more orbits trap.

Owing to the snapshot nature of observations in the real universe, understanding the transfer of angular momentum via observations is all but intractable. Predictions of angular momentum transport have largely been based on analytic theory \citep{tremaine84,weinberg85}, which suffers from necessary idealisations (e.g. fixed bar pattern speeds or small amplitude perturbations). Analytic theory can only predict evolution in the linear regime (owing to its lack of time dependence, see \citealt{weinberg04} for some progress). While the ensuing decades have both confirmed the applicability of linear theory in many cases, simulations have consistently demonstrated a panoply of mechanisms that are not clearly interpreted via linear theory, such as bar buckling instabilities and bar destruction. Hence, a modern picture of bar dynamics must move beyond analytic theory limitations to explore and explain the rich non-linear processes that occur in real barred galaxy evolution. The primary culprit in the breakdown of the analytic linear theory is likely the assumed validity of the averaging theorem, i.e. time averaging over many orbital times, as compared to the real universe, where orbits (stars) may complete only tens of rotation periods during a Hubble time, and the constancy of the gravitational potential.  The potential may change secularly at a rate comparable to the orbital period of individual stars, not to mention other changing quantities in the galaxy like mass accretion or satellite harassment. Here, we investigate one such non-linear phenomenon: the trapping of orbits into patterns.

These realistic complexities of angular momentum transport motivate $n$-body simulations. However, $n$-body simulations are fraught with their own uncertainties (numerical precision, prescriptive evolution for processes below resolution limits), and oftentimes have little ability to definitively implicate the mechanisms observed in analytic calculations as the cause of the $n$-body features. A mechanistic understanding of this important fundamental physics will help build confidence in the physical validity of numerical models as large simulations continue to be the primary means to inform observations.

In \citetalias{petersen16a}, we presented a new manifestation of the standard dynamical mechanism that forms stellar bars, the trapping of the dark matter halo component, which we call the shadow bar, and detailed its impact on secular evolution. We extend the rudimentary angular momentum transfer analysis presented in \citetalias{petersen16a} in this paper, providing an analysis of angular momentum exchange between the trapped and untrapped responses to the bar in both the stellar and dark matter components. We analyse how the simulations presented in this paper may be connected with the analytic framework, and where the comparison of analytic and numerical interpretations falls short. A companion paper, \citet[][hereafter Paper I]{petersen18a}, describes the underlying orbital structure and the importance of different potential features, primarily $m\seq4$, for barred galaxy evolution. \citetalias{petersen18a} also demonstrates a method to decompose sets of orbit trajectories into trapped and non-trapped orbits.  A second companion paper, \citet[][hereafter Paper III]{petersen18c}, describes a harmonic analysis interpretation of barred galaxy models. Together, these works provide a detailed mechanistic understanding for a wide range of evolutionary scenarios.

Using an $n$-body simulation, \cite{athanassoula02} integrated orbits in a fixed potential using a procedure similar to that of \citetalias{petersen18a}, demonstrating in specialised cases that the slowdown of the bar was the result of a loss of angular momentum from orbits near resonances in the inner disc. The angular momentum was absorbed by orbits near the same resonances in the halo, with possible observable effects for breaking halo axisymmetry with an induced wake \citep{petersen16b}. In this paper, we extend the fixed potential approach to a fully self-consistent simulation and analyse the transfer of angular momentum that is a consequence of temporal evolution. This is an underexplored aspect of barred galaxy model evolution. We explicitly show how angular momentum is transported in two simulated barred galaxies. We describe the torque mechanisms by accounting for the changes in specific angular momentum and by a systematic study of the applied forces for each individual orbit. We connect the mechanisms to observables through studying the gross properties of the trapped orbits, their phase-space distribution, the change in angular momentum of individual orbits, and the torques applied by different dynamically-relevant ensembles of material.

This paper is organised as follows. We describe the initial conditions and $n$-body integration technique in Section~\ref{subsec:integration} and relevant analysis tools in Section~\ref{subsec:baridentification}. The gross properties of the bar and its evolution are described in Section~\ref{sec:grossproperties}, including an observationally-motivated look at the simulations in Section~\ref{subsec:phasespace}. We track the angular momentum transfer for orbits in Section~\ref{sec:angmom}, first by looking at the change in angular momentum for individual particles, then by looking at the applied torque. Section~\ref{sec:discussion} contextualises the results and its implications for galaxy evolution and discusses a potential application to integral field unit (IFU) observations. We conclude in Section~\ref{sec:conclusion}.

\begin{figure} \centering \includegraphics[width=3.5in]{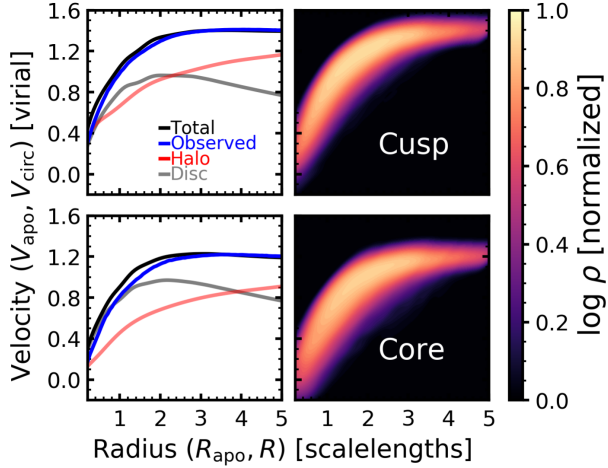} \caption{Initial conditions for the two simulations. The upper (lower) two panels describe the cusp (core) initial conditions.  We plot the decomposition of the circular velocity curve in the left panels, with the disc in grey, the contribution from the halo in red, the total in black, and the `observed' circular velocity in blue (quantified as the azimuthally-averaged tangential velocity computed directly). In the right panels, we plot the log density computed directly from the particles using their apocenter radius, $\rapo$, and corresponding tangential velocity at apocenter, $\vapo$. We normalise the density such that the peak of the density distribution has a value of 1. \label{fig:initialconditions}} \end{figure}

\section{Methods} \label{sec:methods}

\subsection{$n$-body Simulations} \label{subsec:integration}

The simulations explored here were described in detail in \citetalias{petersen18a}. We briefly review the realisation of the initial conditions and integration. We study two model galaxies, called the `cusp' and the `core' simulations for the shape of the dark matter halo profile, described below. Both model galaxies have a stellar disc ($N_{\rm disc}\seq10^6$) and a dark matter halo ($N_{\rm halo}\seq10^7$, $N_{\rm halo,eff}\seq10^9$; see below for details).

We use a $\Lambda$CDM-motivated Navarro-Frenk-White \cite{navarro97} halo model with $\rho_{\rm halo}(r) \propto (r+r_c)^{-1}(r+r_s)^{-2}$. The quantity $r_s$ is the scale radius, defined by the concentration parameter $c\seq R_{\rm vir}/r_s\seq20$, and $R_{\rm vir}$ is the virial radius of the halo. We include an error function truncation outside of $2R_{\rm vir}$ to give finite mass, $\rho_{\rm halo, trunc}(r) = \rho_{\rm halo}(r)\left[\frac{1}{2}-\frac{1}{2}\left({\rm erf}\left[(r-r_{\rm trunc})/w_{\rm trunc}\right]\right)\right]$, where $r_{\rm trunc}=2R_{\rm vir}$ and $w_{\rm trunc}=0.3R_{\rm vir}$. The difference between the two sets of initial conditions is the value of $r_c$, the halo core radius. The galaxy model with $r_c\seq0$ has a $\rho\propto r^{-1}$ cusp that persists to the centre of the model and is, therefore, called the `cusp' simulation. The galaxy model with $r_c\seq0.02$ has a halo density profile that flattens to $\rho\propto$ a constant value towards the centre of the model, resulting in a harmonic core. We call this model the `core' simulation.

We scale all units to so-called virial units, $R_{\rm vir}\seq1$ and $M_{\rm vir}\seq1$, which describe the mass and radius of the halo in a $\Lambda$CDM cosmology with $G\seq1$. In physical units, if we take the halo mass to be that of the Milky Way (MW), $M_{\rm vir}\seq1.3\times10^{12} \msun$ \citep{blandhawthorn16}, then the unit of time is 2.0 Gyr, the unit of length is 300 kpc, and the unit of velocity is 150 \kms. The disc particles all have equal mass, $m_{\rm disc}\seq2.5\times10^{-8}$. We assign the halo particles masses based on their initial radii, following a number density $n_{\rm halo}\propto r^\alpha$, where $\alpha\seq-2.5$. Using this `multimass' scheme, the halo particle mass at radii $r<2a$ ($a$ is the disc scale length; see below) are equal to or smaller than the mass of disc particles. Therefore, near the stellar disc, our multimass halo is equivalent to a halo with $N_{\rm halo,eff}\seq10^9$.

The stellar disc density is modelled as \begin{equation} \rho_{\rm disc}(r,z) = \frac{M_{\rm disc}}{2\pi a^2}\exp\left(\frac{-r}{a}\right){\rm sech}^2\left(\frac{z}{z_0}\right) \label{eq:discdist} \end{equation} where $M_{\rm disc}\seq0.025M_{\rm vir}$ is the total mass of the disc, $a\seq0.01R_{\rm vir}$ is the scale length, and $z_0\seq0.001R_{\rm vir}=0.1a$ is the scale height for the isothermal vertical distribution, described by the ${\rm sech}^2$ distribution. We choose the velocities in the halo from the energy distribution determined by an Eddington inversion of the phase-space distribution function, and in the disc by solving the Jeans equations. The Toomre $Q$ parameter sets the radial velocity dispersion in the disc: $\sigma_r^2(r)\seq3.36\Sigma(r)Q\kappa(r)^{-1}$ where $\Sigma(r)$ is the surface density and $\kappa$ is the radial frequency. We take $Q=0.9$ to promote rapid bar growth. We set the velocity ellipsoid to be axisymmetric, $\sigma_r\seq\sigma_\phi$.

We do not include a bulge in the present models. For bulges less than 10 per cent of the stellar disc mass, a preliminary simulation suggests that the bulge does not play an appreciable role in the evolution described here. Furthermore, a bulge naturally forms during the simulations.

The upper left panel of Figure~\ref{fig:initialconditions} shows the total initial circular velocity curve for the cusp simulation calculated from the basis-expansion potential field (black curve) using $v_c^2\seq R\left(d\Phi_{\rm disc}/dR + d\Phi_{\rm halo}/dR\right)$ where $\Phi$ is the potential. We compute the contributions to the potential from the disc and the halo separately, and show these results in grey and red, respectively. We plot the tangential velocity curve measured directly from the particle distribution, $v_{\rm tan} = (x\dot{y} - y\dot{x})/(x^2 + y^2)$, in blue. Owing to noncircular motions, the measured tangential velocity curve and the calculated circular velocity curve do not match; the two curves grow more discrepant as the simulation evolves. The central portions of galaxy rotation curves with apparent non-axisymmetric structures should be regarded with caution for the purposes of interpreting the enclosed mass, i.e. $v_c \ne \sqrt{GM/r}$ when non-axisymmetric structures are present. The upper right panel of Figure~\ref{fig:initialconditions} shows the log density distribution of the initial particle distribution, where the particles have been placed on the $\rapo-\vapo$ plane, where $\rapo$ and $\vapo$ are the radius and speed of the trajectory at apocenter, as discussed in \citetalias{petersen18a}. Since the locations of non-circular trajectories are dominated by their time at apocenter, we characterise orbits by the phase-space quantities at apocenter.  Similarly, the quantities at apocenter better approximate the features of observed orbits.

The lower panels of Figure~\ref{fig:initialconditions} show the same circular velocity curve, decompositions, and log density as in the upper panel, except for the core simulation. Owing to the lower mass from the dark matter halo at these radii, the rotation curve peaks at a lower maximum rotation speed, despite an identical overall halo mass. The initial discs are identical in density and profile between the two simulations.

We integrate orbits using the basis function expansion (BFE) code {\sc exp} (\citealt{weinberg99}; \citetalias{petersen18a}), which creates two orthonormal potential-density basis sets: one for the disc, in cylindrical coordinates, and one for the halo, in spherical coordinates. For each basis, the lowest order potential profile matches that of the component exactly. A more detailed description of the parameters of the simulation can be found in \citetalias{petersen18a}, and we describe the basis selection in \citetalias{petersen18c}. Particles are advanced using a leapfrog integrator. We save each particle position at every time interval $\delta T\seq0.002$, the master timestep. However, we integrate the orbits based on a timestep criteria defined in \citetalias{petersen16a}, allowing the timesteps to decrease in factors of two. The smallest timestep may be up to $2^4$ times smaller than the master timestep. In practice, over 90 per cent of the disc orbits are always integrated at the smallest time step, $\delta T\seq1.25\times10^{-4}$, which is $<0.002$ of a characteristic dynamical time.  At each intermediate timestep below the master timestep, the coefficients are partially accumulated for orbits that participate in the timestep to update the basis. We fully recompute all the basis coefficients at each master timestep. We require such a fine resolution of saved phase spaces to properly analyse the evolutionary phases, and in particular the transition between evolutionary phases. We evolve the simulation for 4.5 time units. When scaled to the MW, this is 9 Gyr of evolution.

\begin{figure} \centering \includegraphics[width=3.5in]{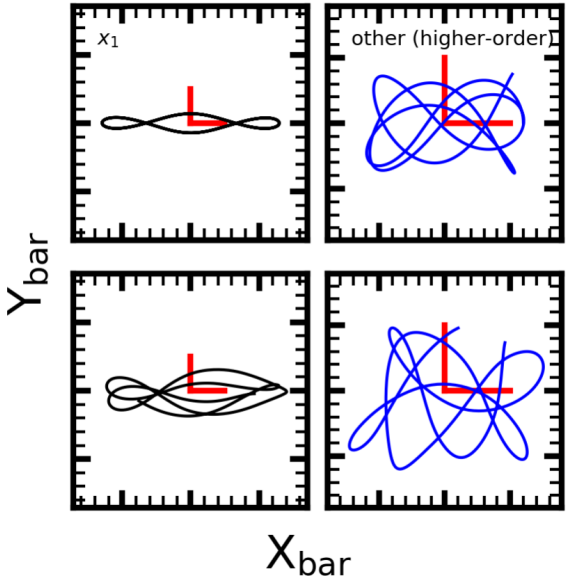} \caption{\label{fig:example_classify} Example classified orbits. The upper panels show theoretical orbit family members drawn from a fixed potential extracted from the cusp simulation at $T=2$, and integrated as in \protect{\citetalias{petersen18a}}. The lower panels show orbits extracted from the self-consistent cusp simulation, highlighting the difficulty of obtaining family membership over small windows of time. The left column shows an $x_1$ orbit. The right column shows an `other' higher-order bar-supporting orbit, following the classification scheme used in this paper.} \end{figure}

\subsection{Bar Identification} \label{subsec:baridentification}

There is no single established method for bar characterisation in the literature. Motivated by observations, one could choose to call the entirety of the observed stellar elongated feature `the bar'. However, this definition assumes that all the orbits in the enclosed region are part of the bar, which we demonstrate to be false. Much analytic theory has shown that the backbone of the bar feature is a particular resonant orbit family ($x_1$ in the parlance of \citealt{contopoulos80}, see \citealt{binney08}), parented by the ILR. We call the orbits whose apsides librate about the bar position in a frame corotating with the bar {\it trapped orbits} and refer to the ensemble of trapped orbits as {\it the bar}. We refer to orbits that linger near the potential minima created by the trapped orbits but that are not librating as {\it dressing orbits}: they `dress' the trapped orbits by residing in the same location in physical space.

We choose to define the bar as the collection of orbits that are trapped into libration. Our technique is theoretically motivated by analytic commensurate orbit analysis \citep[e.g.][]{sellwood93} and resonant evolution \citep[e.g.][]{binney08}, allowing us to more accurately capture the relevant dynamical quantities for secular evolution. For this work, we identify the bar-supporting orbits that librate with the pattern speed of the bar, $\Omega_p$, to compute the potential of the bar, and also to identify the net angular momentum of the bar feature and its role in the transfer of angular momentum throughout the system.

We compute the pattern speed of the bar, $\Omega_p$, from the complex phase of the coefficients of the quadrupole basis functions that are naturally obtained from integration using {\sc exp} (fully described in \citetalias{petersen18a}). The procedure is analogous to Fourier analysis of barred galaxies. As the bar is the strongest quadrupole feature at all times post formation, the quadrupole coefficients primarily describe the bar. From the time series of the quadrupole basis functions, we compute the finite difference at each timestep, providing a measurement of $\Omega_p$. 

 We automate the identification of the commensurate orbits using a $k$-means technique \citep{lloyd82}. We describe the full procedure in \citetalias{petersen18a}. Briefly, the $k$-means technique divides a collection of points into $k$ clusters. In our implementation, we determine the position of all apsides for each orbit. We then transform the $(x,y)$ coordinates for each apsis into the rotating bar frame, $(x_b,y_b)$. At each apsis in the set, we make a series of the 20 nearest apsides in time, using the $(x_b,y_b)$ positions to minimise the separation from $k$ iteratively determined centroids. Once the centroids have been computed, we calculate the phase angle $\phi_k\seq {\rm arctan}\left(y_{{\rm centroid},k}/x_{{\rm centroid},k}\right)$. We define the maximum of the $k$ values of $\phi_k$ to be $\tpos$, the maximum deviation from the bar position angle in the set of 20 consecutive apsides. For determining bar membership, we restrict our analysis to $k\seq2$. For each orbit in the simulation, at each apsis in time, we compute $\tpos$, $\srad$, and $\spara$. Clear groupings emerge at late times when the bar is fully established, from which we empirically calibrate a matched filter to identify known orbit families. We describe the matched filter in detail in \citetalias{petersen18a}.

We classify orbits into two groups: \(x_1\) and \emph{other} bar-supporting orbits. The $x_1$ orbits use the nomenclature consistent with classic orbit studies, e.g. \citet{contopoulos89}. The `other' bar-supporting orbits reinforce the potential of the bar, but are distant enough from commensurate orbits that their family membership cannot be identified. The limits for membership in each family and the details of the classification process are also given in \citetalias{petersen18a}. Once we have associated individual orbits with the bar, a range of physically motivated quantities are available: mass, angular momentum, and length. In later sections, we use these quantities to characterise dynamical processes in the simulation. In Figure~\ref{fig:example_classify}, we show example orbits drawn from the simulation and matched (lower panels) to a classified family identified in a fixed potential orbit calculation centred on the same time interval that is nearby in phase space (upper panels). The left column shows an $x_1$ orbit and the right column shows an `other' bar supporting orbit. The $x_1$ orbit is the only classified family that clearly resembles its parent orbit. However, when using the $k$-means apsis classifier, the other bar orbits are easily identified as having little precession in their apsides over extended time windows. Additionally, we are able to identify a prominent subfamily of the $x_1$ orbit, the bifurcated $x_{1b}$ family. In \citetalias{petersen18a}, we described the classes of orbits available in both models, finding that certain subfamilies of $x_1$ orbits, the bifurcated family $x_{1b}$, in particular, are important for bar growth. We detail the methodology to determine membership in the $x_{1b}$ subfamily in \citetalias{petersen18a}, and refer the interested reader to that paper.

In \citetalias{petersen18a}, we mapped commensurabilities for the cusp and core simulations using two different techniques: a geometric algorithm to find strongly non-circular commensurate orbits in an orbit atlas, and a frequency map derived from the monopole potential to find CR and OLR. Throughout this work, we will show commensurabilities determined via the geometric algorithm as white overlays, and commensurabilities determined via the frequency map as cyan overlays. We refer the reader interested in the determinations of the commensurability structure to \citetalias{petersen18a}, and simply apply the results to this work where applicable. Where possible, we identify (1) bar orbits in the $x_1$ family, which are parented by the ILR, including long period $x_{1l}$ orbits and short-period $x_{1s}$ orbits as well as bifurcated $x_{1b}$ orbits; (2) orbits that exhibit 3:$n$ symmetry, where 3 corresponds to the number of radial oscillations per $n$ rotation periods, in a frame co-rotating with the bar; (3) orbits associated with CR; and (4) orbits associated with OLR.

\section{Simulation Gross Properties} \label{sec:grossproperties}

We describe the phase space distributions of the simulation in the observationally-motivated $\rapo-\vapo$ plane in Section~\ref{subsec:phasespace}. We characterise the bar using a variety of dynamically-informative metrics in Section~\ref{subsec:barproperties}. In Section~\ref{subsec:untrapping}, we discuss and quantify the rate of previously trapped orbits leaving the bar.

We identify three phases of evolution for the simulations, denoted by different phases in the bar lifetime \citepalias{petersen18a}: \begin{enumerate} \item The bar {\it assembles} owing to a local dynamical instability. \item The bar {\it grows} owing to a secular instability. \item The bar evolution practically ceases; a {\it steady state}. The steady-state may be transient or long-lived. \end{enumerate} We use these three phases throughout the paper to understand the mechanisms present during the simulation. Although all bars start with the assembly phase, the other two phases can proceed in any order and can occur more than once. In the simulations presented here, the order after assembly is growth then steady-state for the cusp simulation and steady-state followed by growth for the core simulation \citepalias[see][for details]{petersen18a}. The most well-understood of the three phases is the growth phase. The growth phase is the standard secular evolution phase, well-studied by perturbation theory. We largely find that known secular processes can explain the evolution in the growth phase. The other two phases have been probed by simulations, but still lack a full dynamical explanation. The inability to track dynamical quantities and their changes owing to nonlinear processes has limited the understanding of mechanisms in simulations. We, therefore, attempt to make progress by providing further information about the phase space and gross properties of the bar during these phases.

Overall, we find that simulations constructed with different halos exhibit varied evolutionary timescales and patterns. However, during the identified epochs of evolution (assembly, growth, and steady state), the governing physics in each simulation is the same.

\subsection{Phase-Space Distributions} \label{subsec:phasespace}

\begin{figure*}
  \centering \includegraphics[width=5.0in]{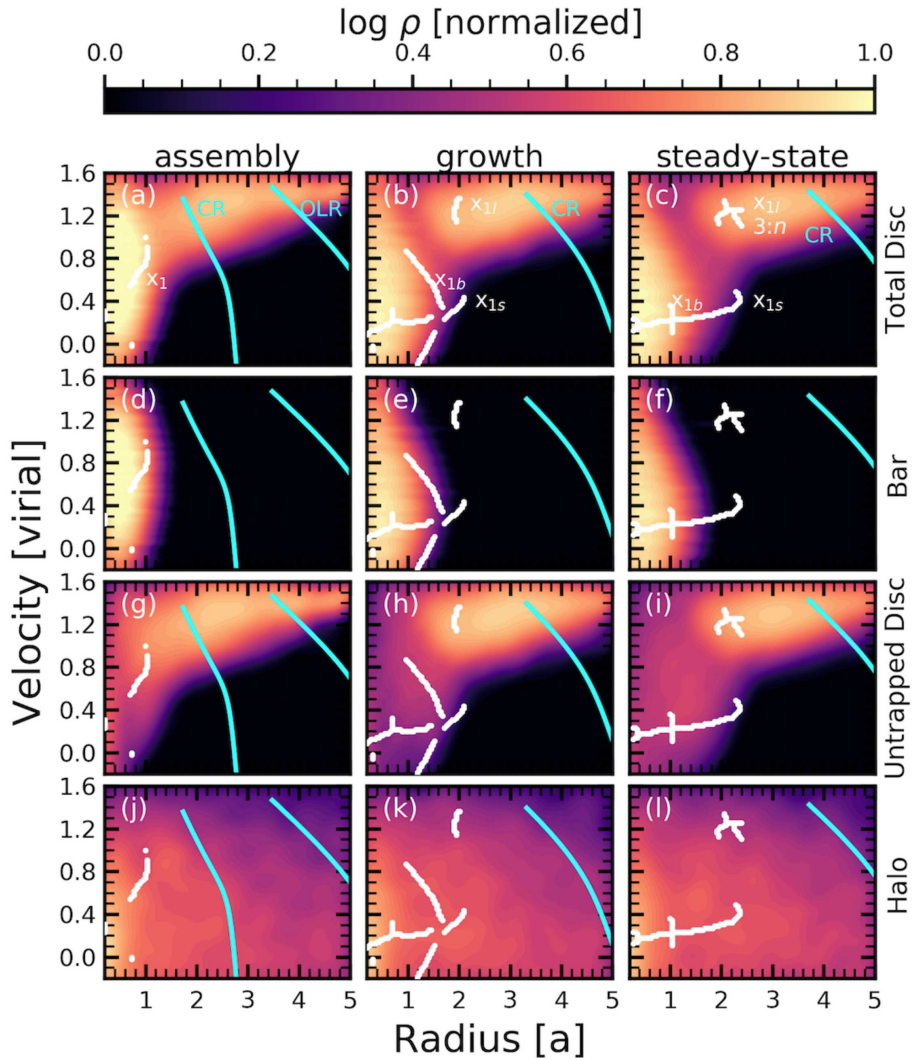} \caption{Log density for the cusp simulation in the $\rapo-\vapo$ plane, normalised to the highest log density in the total disc (such that each panel has the same stretch), for four different ensembles of particles (total disc, bar, untrapped disc, and halo, top to bottom). The columns correspond to the observed phases of evolution, assembly (left column), growth (middle column), and steady-state (right column). For each column, we overlay the commensurabilities calculated and coloured using the methodology described in Section~\ref{subsec:baridentification} and \protect{\citetalias{petersen18a}}. The main commensurate families are labelled in panels a, b, and c, and apply to every row in the corresponding column. \label{fig:cusp_density}} \end{figure*}

\begin{figure*} \centering \includegraphics[width=5.0in]{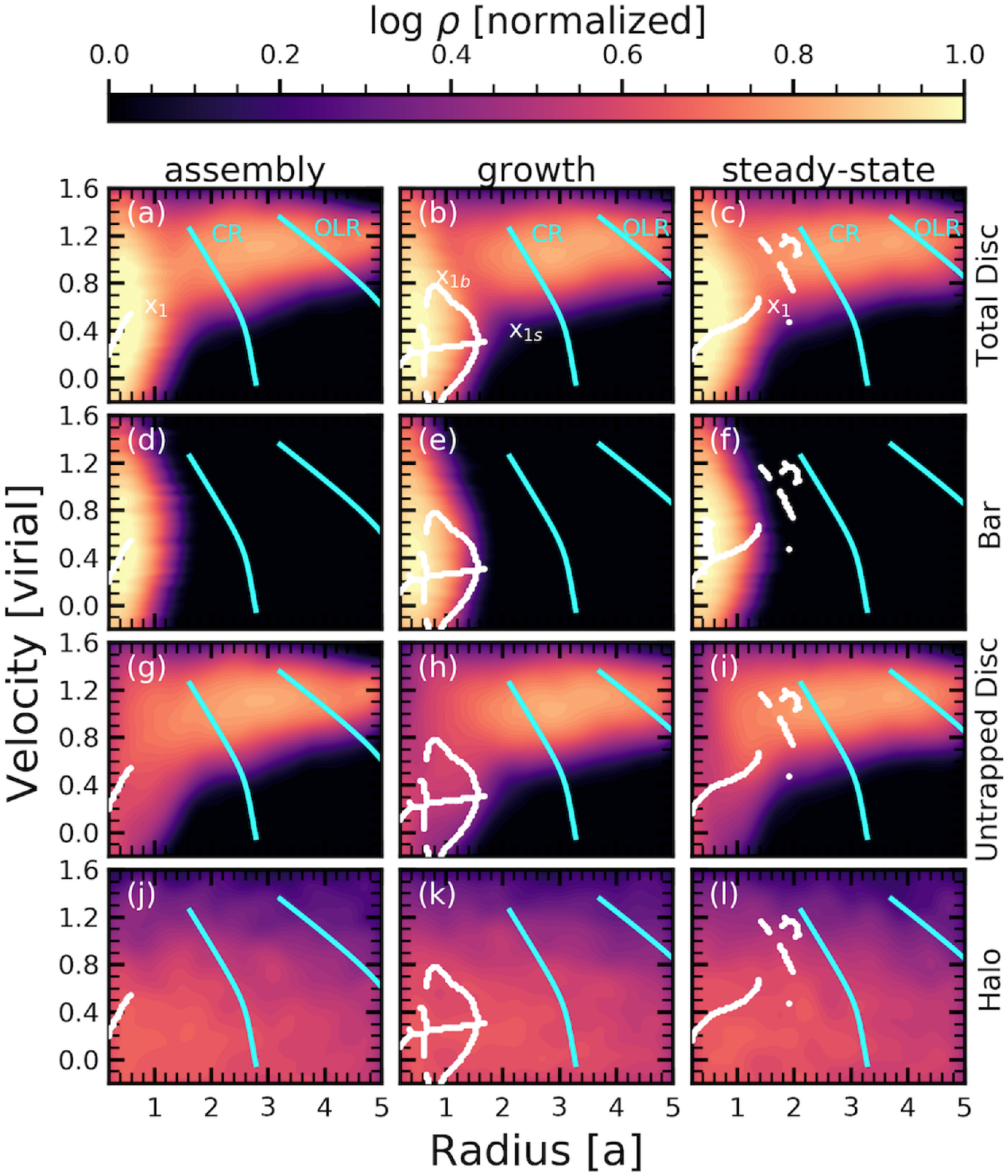} \caption{Same as Figure~\ref{fig:cusp_density}, but for the core simulation. \label{fig:core_density}} \end{figure*}

In Figure~\ref{fig:cusp_density}, we show the normalised log phase-space density from the cusp simulation in the $\rapo-\vapo$ plane for three components (bar, untrapped disc, and halo), as well as for the total disc (a combination of the bar and untrapped disc) during the three evolutionary phases. The normalisation is the same across all panels in the figure, so that relative densities can be compared. We compute the density distributions using instantaneous values drawn from the simulation at the centre of each phase. Therefore, this is not the true $\rapo-\vapo$ space distribution. However, since the orbital positions linger near $\rapo$ for a larger fraction of their trajectories, this distribution of radius and velocity is not too different from the true $\rapo-\vapo$ distribution. We use the instantaneous radius and instantaneous planar tangential velocity $v_t = (x\dot{y} - y\dot{x})/(x^2+y^2)$ to place orbits on the radius-velocity plane. For the rest of the work, when we refer to the radius-velocity plane, the velocity is the planar tangential velocity.

The upper row of Figure~\ref{fig:cusp_density} shows the density distribution for the total disc at the three identified phases (panels a, b, and c). The observed changes between each phase, and in particular the appearance of a valley in phase-space density between the bar and the bulk of the untrapped disc material, may provide observational hints of when a bar is dynamically young and still assembling. Additionally, we overlay the commensurability structure for the three phases as mapped in \citetalias{petersen18a}. The difference in orbit families between the assembly phase (panel a) and the growth phase (panel b) are stark, representing the rapid reorganisation of the disc in response to the transfer of angular momentum.

We separate the disc into the trapped bar (panels d, e, and f of Figure~\ref{fig:cusp_density}) and untrapped disc (panels g, h, and i of Figure~\ref{fig:cusp_density}). The bar assembly process, which is ongoing in panels d and g, has not yet drawn a significant fraction of orbits out to a disk scale length $a$ (see the detailed analysis of trapped orbits in Paper I). During the growth phase (panels e and h), the majority of the disc material interior to 1.5$a$ joins the bar, though some disc material remains untrapped at the same radii. Untrapped material persists even during the steady-state phase at similar radii and velocities to that of the bar (panels f and i).\footnote{The presence of untrapped material at radii smaller than a bar length demonstrates the observationally confusing role of dressed orbits (cf. Section~\ref{subsec:baridentification}) in determining bar properties.}

The phase-space density distribution for the steady-state phase (panels c, f, and i) reveals a gap in phase-space density demarcated by the segment (0.5, 1.6) to (2.0, 0.6) in $(\rapo,\vapo)$.  This is especially clear in the decomposition of the full distribution into bar and untrapped components. This phase-space density minimum in radius-velocity space may be used as an indicator of an evolved barred galaxy.  Further, the length of the bar is limited by the $x_{1b}$ track identified in \citetalias{petersen18a}, rather than the extent of the $x_1$ track, showing that the $x_{1b}$ family plays an important role in the dynamics of barred galaxies. We label the bifurcation of the $x_1$ family, the $x_{1b}$ family, in panels b and c of Figure~\ref{fig:cusp_density} \citepalias[see][for a detailed description of the orbit families]{petersen18a}.

We plot the same quantities for the core simulation in Figure~\ref{fig:core_density}. As we already mentioned, a steady-state evolution phase precedes the growth phase in this model, but to facilitate the comparison of mechanisms that drive both phases, we show the phases in the same order as in Figure~\ref{fig:cusp_density}. We note many similarities between the two simulations, indicative of the similarity in mechanisms during the different phases. During the steady-state phase (panel c), which occurs {\it before} the growth phase in this simulation (panel b), the average velocity of bar particles is the same as during the assembly phase. As in the cusp simulation, during the growth phase (panel b), when the $x_{1b}$ orbit family is present, the length of the bar coincides with the location of the bifurcation. We mark the location of the $x_{1b}$ family in panel b of Figure~\ref{fig:core_density} (the $x_{1b}$ family does not exist in the other identified phases). As in Figure~\ref{fig:cusp_density}, we mark the approximate location of CR. The assembly and growth of the bar again occurs when CR is located in a region of high orbit density (left and centre columns). However, unlike in the cusp simulation, CR is located in a region of high orbit density during the steady-state phase as well (right column). This suggests that the location of CR is not indicative of the state or phase of the simulation, and potentially calls into question the ubiquity of using CR to interpret observations of barred galaxies. We also delineate OLR, which is present within the five scale length cutoff of our panels. The total disc presents a weaker minimum in phase-space density between the bar and the disc in $\rapo-\vapo$ space (panel c) than in the cusp simulation, consistent with our understanding that the steady-state phase in the core simulation is a transient phase.

In general, the simulations behave in the same manner during the corresponding evolutionary epochs, e.g. the growth phase in the cusp simulation exhibits the same characteristics as the growth phase in the core simulation. Though the epochs happen in a different order, we shall see below that these similarities correspond to similar governing physics during each corresponding epoch.

\begin{figure} \centering \includegraphics[width=3.0in]{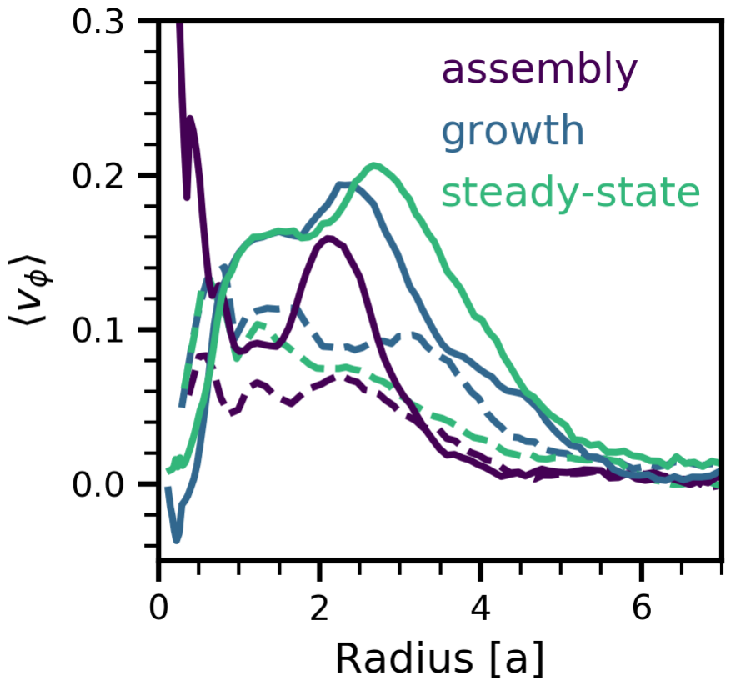} \caption{Mean tangential velocity, $\langle v_\phi\rangle$, as a function of planar radius, $r=\sqrt{x^2+y^2}$, for the dark matter halo in the cusp (solid lines) and core (dashed lines) models. For each simulation we plot the three phases, colour-coded as shown. The cusp simulation accepts more angular momentum than the core simulation, and thus $\langle v_\phi\rangle$ is always higher at all disc radii for the corresponding times. As the simulation progresses, more angular momentum is deposited and the mean tangential velocity both increases and moves to larger radii. \label{fig:mean_lz}} \end{figure}

While not evident in Figures~\ref{fig:cusp_density} and \ref{fig:core_density}, the distributions in the halo (panels j, k, and l) skew modestly positive as compared to an isotropic distribution with $\langle v_\phi\rangle=0$. Further, a clear but subtle difference between the cusp and core simulations is evident in the density distributions. To illustrate this, we show the mean tangential velocity $\langle v_\phi \rangle$ as a function of planar radius $r=\sqrt{x^2+y^2}$ in Figure~\ref{fig:mean_lz}. The run of $\langle v_\phi \rangle$ is positive within approximately four disc scale lengths, indicative of deposited angular momentum (the flow of which we detail below) and that the assembly of the bar dynamically alters the inner halo \citep{weinberg02}. However, as the cored halo has less halo mass in its interior, the dynamical effects that are observed in the cusped halo are less pronounced for this model: $\langle v_\phi\rangle$ is less positive at all times. The strongest peak in the $\langle v_\phi \rangle$ distribution corresponds to the deposition of angular momentum at corotation in the halo. Corotation consistently moves outward in radius in both models as the simulation evolves (see, e.g., the commensurability lines in Figures~\ref{fig:cusp_density} and \ref{fig:core_density}). Other peaks in $\langle v_\phi\rangle$ correspond to different resonances, including the ILR, located at approximately one scale length for all curves.

\subsection{Bar Properties} \label{subsec:barproperties}

\begin{figure*} \centering \includegraphics[width=5.8in]{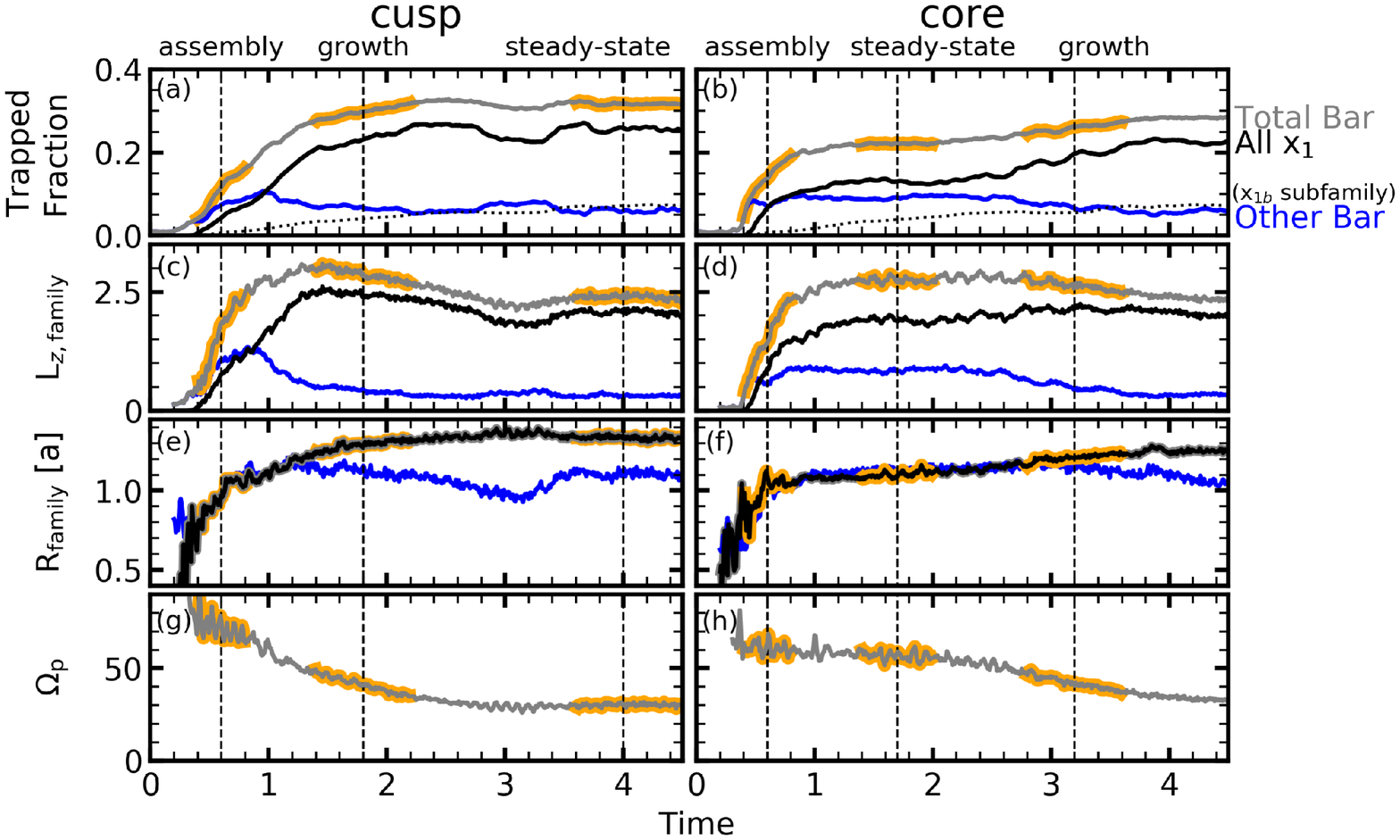} \caption{Bar property evolution in the cusp simulation (left panels) and core simulation (right panels). Panels a and b: Trapped fraction of disc orbits for two classifications: $x_1$ (black) and other bar orbits (blue). We additionally show the $x_{1b}$ subfamily as a dotted line. The sum of the trapped fractions, comprising the total bar fraction, is shown as a grey line. Three periods of bar evolution are defined: assembly, growth, and steady-state, labelled above the top panel. Panels c and d: Total angular momentum in each component listed in the upper panel. Panels e and f: 99$^{\rm th}$ percentile radius for all members of each family. Panels g and h: The pattern speed, $\Omega_p$, computed for the total bar (Section~\protect{\ref{subsec:baridentification}}). \label{fig:cusp_angmom} } \end{figure*}

Using our $k$-means and geometric orbit classification tools, we characterise the bar by orbit family with mass $M_{\rm family}$, angular momentum $L_{{\rm z, family}}$, and the radius that encloses 99 per cent of the orbit apocentres in the family\footnote{For the bar-dominant $x_1$ family, we will refer to the radius enclosing 99 per cent of the $x_1$ orbits as the length of the bar.}, $R_{\rm family}\equiv R_{99}$. Figure~\ref{fig:cusp_angmom} shows the evolution of these gross properties for the important bar orbital families in the simulations: $x_1$ and the `other' bar orbits. Each of the quantities is dynamically meaningful. We also show the pattern speed for the total bar, computed from the coefficient time series (Section~\ref{subsec:baridentification}). We highlight relevant points in the evolution for further dynamical investigation in Section~\ref{sec:angmom}. Generally speaking, changes to the sign of the slope in each quantity tend to signal a new era in evolution, though all are symptoms rather than underlying causes of the dynamical status---we will discuss the causes in Section~\ref{sec:angmom}. We plot all the quantities in Figure~\ref{fig:cusp_angmom}, where the left column is for the cusp simulation, and the right column is for the core simulation.

\subsubsection{Bar Mass} \label{subsubsec:barmass}

The bar mass is the mass of the particles trapped into libration by the bar. We measure two classes of trapped orbits: orbits that are the backbone of the bar--the $x_1$ family and the $x_{1b}$ subfamily--and orbits that are trapped into the bar as higher-order families, called `other' bar supporting orbits. The sum of the $x_1$ family and `other' bar supporting orbits makes up what we call the total trapped bar orbits. We use the trapped fraction of the bar as the main descriptor for the evolutionary phases that occur in the simulation. We shall see below that many untrapped orbits reside in the same physical regions of a galaxy. These untrapped orbits would be considered part of the bar in a standard ellipse-fit based analysis, though they do not participate in the dynamics in the same manner as trapped orbits, remaining distinct from the angular momentum transfer of the trapped bar orbits. This is likely a significant source of uncertainty in determining the masses of bars in observations. In our simulations, the dressing orbits may be up to 10 per cent of the total disc mass, or 50 per cent of the total bar mass.

In the panels a and b of Figure~\ref{fig:cusp_angmom}, we show the trapped fraction as a function of time for the $x_1$ family (solid black), including the $x_{1b}$ subfamily (dotted black) and the other bar supporting orbits (solid blue). We sum the $x_1$ and other bar supporting orbits to make the solid grey line, the total trapped fraction of the bar. These panels clearly demonstrate the three evolutionary phases of the bar in each simulation.

In the cusp simulation the evolution proceeds in the order listed above: assembly, then growth, then steady-state (panel a). The core simulation exhibits all three modes (panel b), but upon conclusion of the assembly phase, the galaxy is in an apparent steady-state phase of evolution, where evolution progresses slowly. After some evolution of the potential, new families of stable orbits, principally the $x_{1b}$ subfamily \citepalias[see][]{petersen18a}, appear and the bar begins to grow again. Thus, in panel b, we see an assembly phase followed by a transient steady-state phase, followed by a growth phase. The simulation reaches a second stable steady-state phase at the end of the simulation. The mechanisms are the same as in the steady-state phase of the cusp simulation, as discussed below. In both simulations, an increase in $x_{1b}$ orbits corresponds to a decrease in other bar orbits.

As we observe that the evolution of the two simulations proceed differently, we seek other explanations for why each model behaves as observed. This leads us to look at other metrics to understand the physics that governs the different evolutionary phases, and why the models behave differently after a bar has assembled.

\subsubsection{Bar angular momentum} \label{subsubsec:barangmom}

Panels c and d of Figure~\ref{fig:cusp_angmom} show the angular momentum that resides in each of the three tracked families from the upper panel. We compute the angular momentum in the inertial frame. In the cusp simulation (panel c), the three evolutionary phases of the bar have well-defined trends in their total angular momentum. During the assembly phase, the bar rapidly accumulates angular momentum, as the mass of the bar increases. The increase in angular momentum from the growth of the bar by orbit trapping exceeds the loss of angular momentum by secular evolution.  Therefore, at the same time, the specific angular momentum, or average angular momentum per bar particle, decreases. After the assembly phase concludes, the bar {\it loses} total angular momentum, even as it grows in mass, during the growth phase. This inflection point in the total $\Lz$ is reflected in the structure and evolution of the torque that is applied, as we will discuss below. During the steady-state phase, the bar does not add or subtract angular momentum, owing to a lack of low-order commensurabilities or families to provide a secular coupling and/or the lack of available phase-space in the halo or outer disc to accept angular momentum. \citetalias{petersen18a} describes orbit families that may be readily torqued, and \citetalias{petersen16a} described the implications of reducing the available phase-space for angular momentum acceptance via secular evolution. The core simulation (panel d) reveals the same behaviour as the cusp simulation, for the same evolutionary phase.

The angular momentum grows during the assembly phase, decreases during the growth phase, and increases only slightly during the steady-state phase. The total angular momentum of the bar versus time is the most sensitive indicator of the evolutionary status in both models, as the sign of the slope changes between the assembly and growth phases, and does not change during the steady-state phase.

\subsubsection{Bar length} \label{subsubsec:barlength}

We define the length of the bar to be the radius that encloses 99 per cent of the trapped orbit's apocentres, $R_{99}$. We eliminate the largest 1 per cent to be consistent with our estimate for contamination in the trapping metric, see \citepalias{petersen18a}. This quantity represents the outer radial boundary of trapped orbits. Surface density plots are biased by a relatively small number of orbits that linger near the end of the bar; so-called `dressing' orbits that we will discuss below and in detail in \citetalias{petersen18c}. The $x_1$ family has a longer length than the other bar orbits, with the 99th percentile reaching $R_{99}\approx1.5a$ during the steady-state phase of bar evolution in the cusp model, panel e. $R_{99}$ is an observable quantity given sufficient velocity information \citep[we present a method in][]{petersen18cc}, and is more robust as a bar-length estimator than isophote fitting to low-amplitude noncircular variations. In the cusp simulation (panel e), the length of the $x_1$ family increases rapidly during the assembly phase, continues increasing during the growth phase, and does not change during the steady-state phase. In the core simulation (panel f), we see the same behaviour in the assembly and growth phases, but during the steady-state phase, the $x_1$ orbits increase weakly in length. 

In the traditional view of bar-induced secular evolution, orbits trapped in the bar transfer $\Lz$ to the outer disc and elongate in response while the bar rotates as a rigid body. By inspecting individual orbits, we see that the bar lengthens by trapping near orbits at larger radii. We are careful to point out that the lengthening of bars does not mean that individual orbits change their radial extent; rather, new orbits with larger apocentres join the bar, causing the bar to appear longer. The addition of orbits at the end of the bar was shown in \citetalias{petersen18a}, where the commensurability skeleton traced ILR, and identified that maximal $x_1$ orbits continue to move to larger $\rapo$ throughout the simulation. The $x_{1b}$ orbits are most responsible for bar growth and their position in $\rapo-\vapo$ space limit the size of the bar.

\subsubsection{Bar pattern speed and summary} \label{subsubsec:barpatternspeed}

Panels g and h of Figure~\ref{fig:cusp_angmom} shows $\Omega_p$. In the cusp simulation, $\Omega_p$ decreases during both the assembly and growth phases before becoming constant during the steady-state phase (panel g of Figure~\ref{fig:cusp_angmom}). For the core simulation, $\Omega_p$ is qualitatively similar to the cusp simulation during the assembly and growth phases (panel h). 

In summary, the gross properties of the bar are one way to compare the simulations with one another, and readily reveal the different phases of bar evolution. However, they are also difficult to obtain through observations as they are a product of several underlying quantities for individual orbits, which cannot be observationally resolved. We report and discuss them here in the hopes that the metrics could be used to compare different simulations to help understand the variety of physical mechanisms in the diversity of published simulation results.

\subsection{Untrapping} \label{subsec:untrapping}

\begin{figure} \centering \includegraphics[width=3.4in]{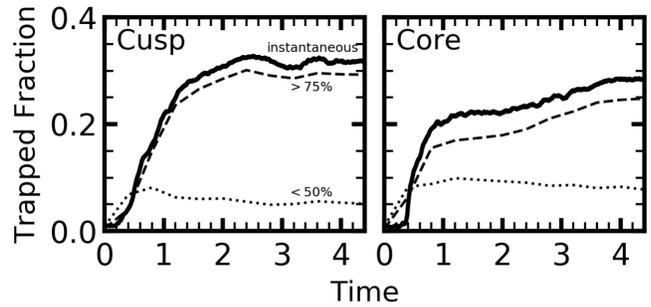} \caption{Bar trapped fraction of disk material versus time for the cusp simulation (left panel) and core simulation (right panel). The solid black line shows the instantaneous trapped fraction (cf. Figure~\protect{\ref{fig:cusp_angmom}}). The dashed line shows the instantaneous measure of orbits that are trapped for at least 75 per cent of each bar rotation. Owing to smoothing over $\Delta T=0.2$ windows, the early time evolution in the core simulation shows this fraction to be larger than the instantaneous fraction. The dotted line shows the instantaneous measure of orbits that are trapped for less than 50 per cent of the time window. \label{fig:untrapping}} \end{figure}

Formally, the secular capture of orbits is a two-way process, where the release of orbits from resonance can also occur \citep{binney08,daniel15}. Our precise determination of membership in the bar at any given time enables us to make quantitative statements about the rate that orbits move from libration to rotation or {\it untrap}. Quantifying the untrapping has implications for observations of stars that are in the vicinity of the bar in the MW, and whether stars born in the bar may be found outside of the bar at later times.

In panels a and b of Figure~\ref{fig:cusp_angmom}, we presented a simple total of the trapped orbits instantaneously at a given time in the simulation. However, this panel does not provide any information about whether the particles that become trapped remain trapped. To address this question, we define the fraction of time an orbit remains trapped in a chosen time window, $\zeta$. We choose to define the quality of orbit trapping using the terms {\it fully trapped} and {\it loosely trapped}. Quantitatively, fully trapped orbits are those that, over a bar period ($\Delta T=0.2$), are associated with a trapped family for $>$75 per cent of the window. Loosely trapped orbits are those that over a given time window are associated with a trapped family for $<$50 per cent. Qualitatively, we do not consider family switching to be untrapping, so an orbit that goes from $x_1$ to an `other' bar supporting orbit or vice versa would not be considered untrapped.

Figure~\ref{fig:untrapping} shows the fraction of disk orbits that are fully trapped (dashed lines) and loosely trapped (dotted lines), as compared to the instantaneous trapped fraction (solid lines) from Figure~\ref{fig:cusp_angmom}.  The left panel shows the bar trapped fraction for the cusp simulation, and the right panel shows the core simulation. The figure demonstrates that approximately 6 per cent of the total disc ($\approx$10 per cent of the instantaneously trapped bar) is loosely trapped in the cusp simulation, and 8 per cent ($\approx$20 per cent of the instantaneously trapped bar) in the core simulation. At any given time, these percentages of orbits join and leave the bar. However, the instantaneous trapped line, shown in black, suggests that the population of orbits that are loosely trapped remains roughly constant with time. However, it is not as simple as a one-way channel, e.g. the orbits are not guaranteed to be loosely trapped before becoming fully trapped. In general, orbits rapidly become fully trapped when they join the bar, and a separate population of orbits that were previously fully trapped into the bar become a loosely trapped population. In terms of physical location, the loosely trapped orbits are located near the end of the bar, where the commensurability density is high. The core simulation always has a higher fraction of loosely trapped orbits relative to the instantaneously trapped orbits compared to the cusp simulation.

While the simulations show only subtle differences, our measurements confirm that trapping orbits into the bar feature is not a mono-directional process. One cannot assume that the orbits observed to be `instantaneously' (or over some short time window) trapped into the bar will remain a part of the bar.

\section{The Angular Momentum Economy} \label{sec:angmom}

We construct a phenomenological picture of the transfer of angular momentum between the stellar bar, untrapped disc, and dark matter halo by partitioning orbits into distinct ensembles and studying the couples between each separately. Each is connected to the other; determining the magnitude of the exchanges connecting the components sheds light on the process of angular momentum transfer in galaxies. Throughout both the dynamical and secular instability phases, angular momentum is rearranged, primarily between the outer disc and the dark matter halo, catalysed by the bar. The dark matter has a resultant wake\footnote{As does the disc outside of a bar radius.}, indicative of the transfer of angular momentum \citep{weinberg85}. The cusp and core simulations are more similar than different in their angular momentum transfer mechanisms. Therefore, we describe the transfer between the components for the cusp case and detail the differences in the simulations where applicable. The evolution of the disc density profile results from a redistribution of energy and angular momentum. We explain the features in the $\rapo-\vapo$ plane as this facilitates comparison to observational metrics.

As in measurements of the bar (Section~\ref{sec:grossproperties}), the simulations show similar physical signatures of angular momentum transfer during the same phases, particularly the growth and steady-state phases.

\subsection{L$_z$ Accounting From Particles} \label{subsec:particletorque}

\begin{figure*} \centering \includegraphics[width=5.0in]{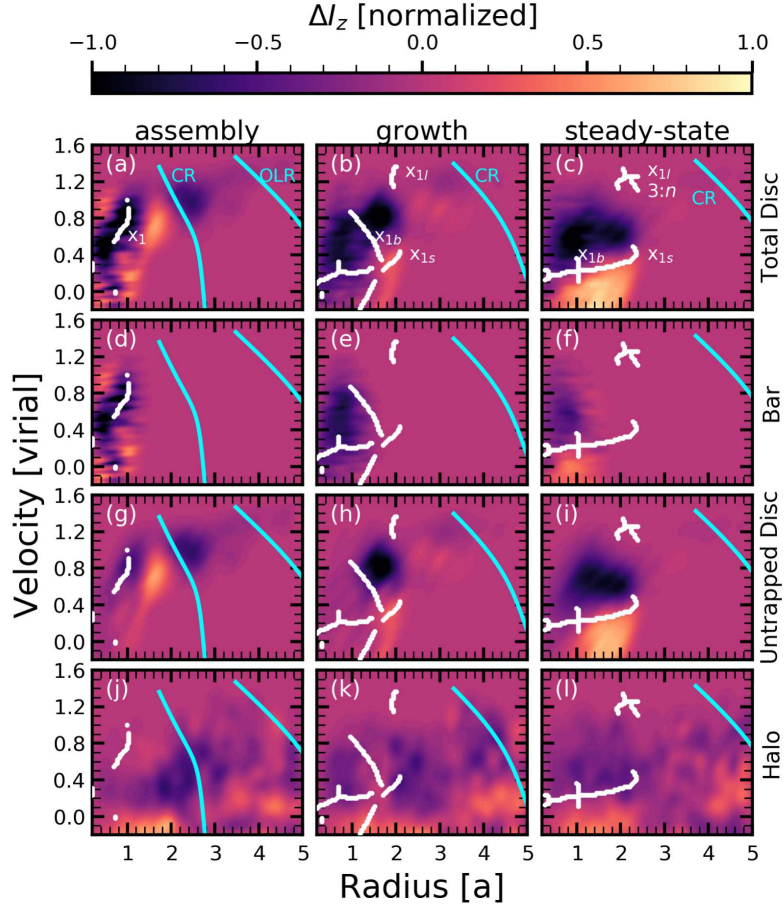} \caption{Transfer of angular momentum in the cusp simulation in $\rapo-\vapo$ space, as computed from finite differencing of ensembles of particles for the cusp simulation. All panels are normalised to the same angular momentum scale, where darker colours mean that the region of $\rapo-\vapo$ space lost angular momentum, and lighter colours mean the region gained angular momentum. Panels a, b, and c: angular momentum change $\Delta \Lz$ for particles in the disc during assembly (left column), growth (middle column), and steady-state (right column).  Panels d, e, and f: same panels a, b, and c except for bar particles. Panels g, h, and i: same as panels a, b, and c, except for the untrapped disc particles. Panels j, k, and l: same as panels a, b, and c, except for the dark matter halo. As in Figure~\ref{fig:cusp_density}, the overlays in white and cyan show the commensurabilities identified at each timestep. \label{fig:cusp_lztransfer}} \end{figure*}

\begin{figure*} \centering \includegraphics[width=5.0in]{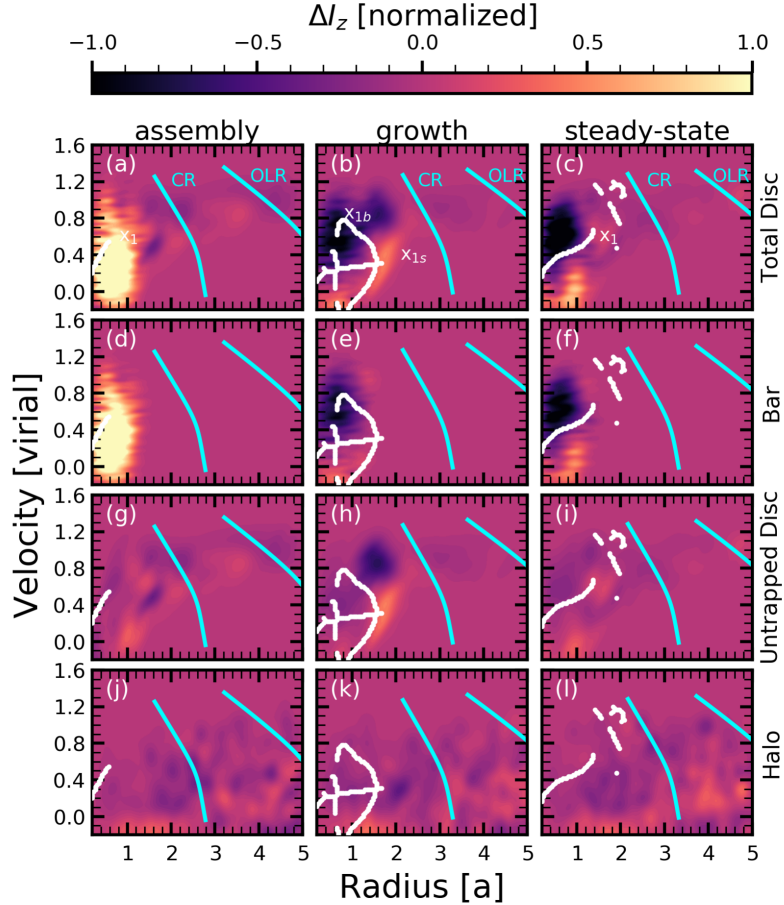} \caption{Similar to Figure~\protect{\ref{fig:cusp_lztransfer}}, except for the transfer of angular momentum in the core simulation. \label{fig:core_lztransfer}} \end{figure*}

\begin{figure*}\centering \includegraphics[width=5.0in]{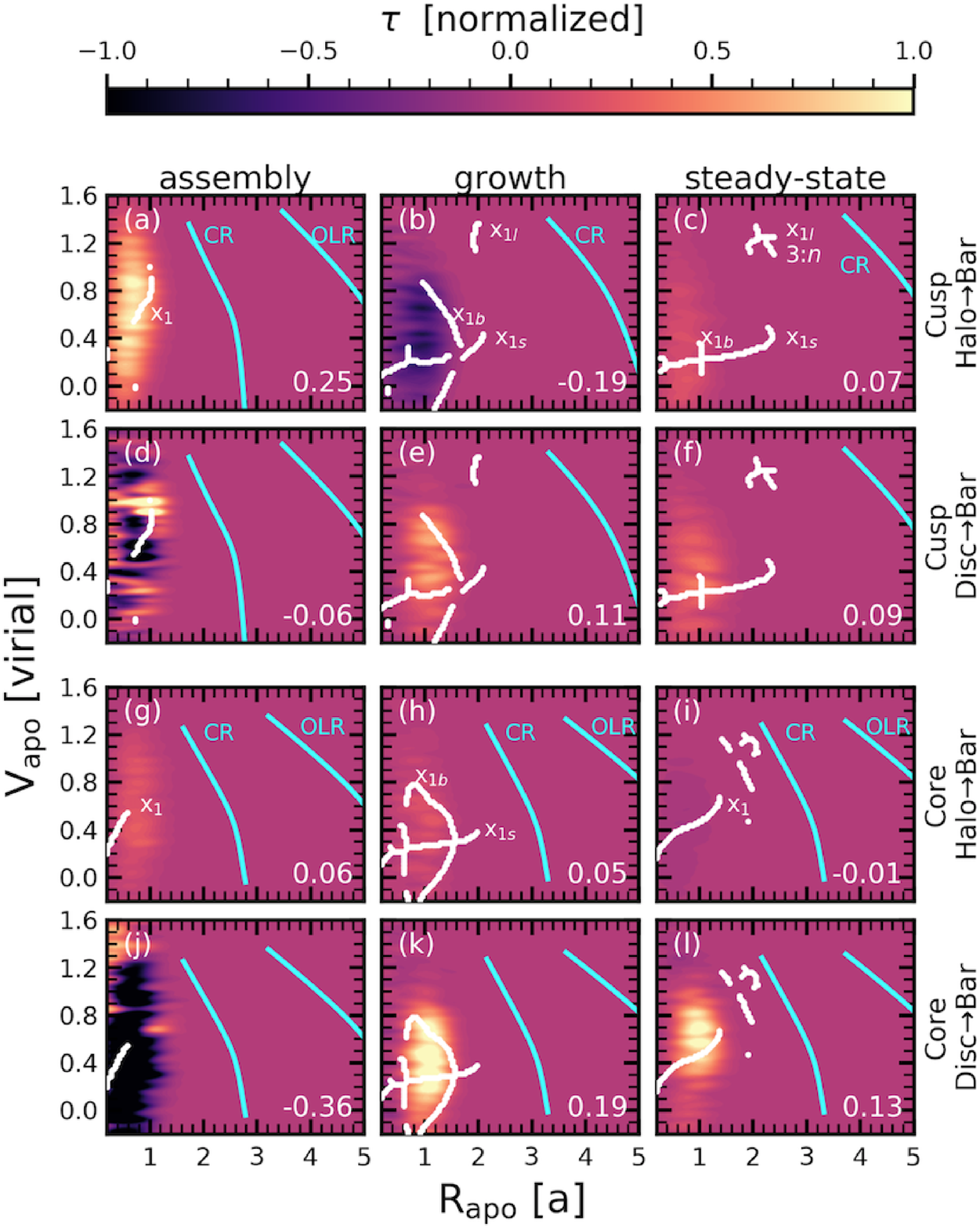} \caption{Instantaneous applied torque in $\rapo-\vapo$ space, as computed from the forces for the cusp simulation (upper block, panels a-f) and core simulation (lower block, panels g-l). Within each block, the upper row is the torque applied by the halo to bar particles in the assembly phase (left column), growth phase (middle column), and steady-state (right column). The lower row in each block (panels d-f and j-l) is the same as the uppermost row (panels a-c and g-i), except it is the torque applied to the bar by the disc particles. Each row is labelled with the simulation and channel to the right. The corresponding commensurability structure has been overlaid on each panel, and commensurabilities have been labelled where relevant. In the lower right corner of each panel, we report the mean torque per particle.\label{fig:net_torque}} \end{figure*}

Individual particles with the same energy and angular momentum respond differently to the bar depending on their angular phase. Therefore, we perform averaging to see the underlying trends that represent the net mean change in conserved quantities. We cancel these first-order phase-dependent variations by (1) using an orbital value for angular momentum derived over a radial period, (2) using a large number of particles, and (3) by placing the particles on the $\rapo-\vapo$ plane where features can be associated with resonances. However, the phase-dependence may persist on timescales significantly longer than our averaging window near resonant degrees of freedom and, therefore, some of the first-order response will still remain. We have verified that the results presented here are not strongly dependent on window size.

For the three windows corresponding to the evolutionary phases defined in Section~\ref{sec:grossproperties} (assembly, growth, steady-state), we compute the radially scaled angular momentum $\Iz = (x\dot{y} - y\dot{x})/(x^2 + y^2)^{1/2}$ averaged over a radial period, which we write as $\langle \Iz \rangle_R$. The scaling allows $\Lz$ at small and large radii to be compared, provided that the rotation curve is relatively flat, as is true in our case. We define the change in angular momentum for a given orbit as $\Delta \Iz \equiv \langle \Iz \rangle_{R, {\rm late}} - \langle \Iz \rangle_{R, {\rm early}}$, where the subscripts `early' and `late' indicate the {\it complete} radial periods closest to the beginning and end of the time windows, respectively. We partition the orbits into bar, disc, and halo ensembles. We described the decomposition in Section~\ref{subsec:baridentification}. The density of the different components in the $\rapo-\vapo$ plane can be found in Figures~\ref{fig:cusp_density} and \ref{fig:core_density}. The ability to decompose the disc into trapped and untrapped particles, even when the orbits reside in the same physical space, enables our angular momentum tracking.

In Figure~\ref{fig:cusp_lztransfer}, we show the transfer of angular momentum computed from the particles for the cusp simulation. As in Figure~\ref{fig:cusp_density}, we examine the angular momentum transfer during the three evolutionary phases (assembly, growth, and steady-state), and for four ensembles of particles (total disc, bar/trapped disc, untrapped disc, halo). Unlike in Figure~\ref{fig:cusp_density}, we compute true values of $\rapo$ and $\vapo$ for the orbits, making the placement of individual orbits in the plane less dispersed and the signals stronger. We again normalise the total angular momentum change, but the scaling is the same between all the panels, so the colour maps may be directly compared. The $\Delta \Iz$ in Figure~\ref{fig:cusp_lztransfer} is the total radial angular momentum, i.e. the sum of all the radially-scaled angular momentum lost by particles at that position in $\rapo-\vapo$ space. A comparison to the specific angular momentum change, e.g. $\Delta \Iz/M_{\rm \rapo, \vapo}$ reveals the total change to be more informative for galaxy evolution.

Panels a, b, and c of Figure~\ref{fig:cusp_lztransfer} show the change in angular momentum for all particles in the disc. The steady-state phase (panel c) is the most straightforward phase to interpret. During this phase, the change in angular momentum is confined to particles within the maximum $x_1$ radius, $R_{{\rm max}~x_1}\le2.5a$. The $x_1$ track splits the orbits that gain and the orbits that lose angular momentum, with the orbits that lose (gain) angular momentum being at higher (lower) velocities than the $x_1$ track. The magnitude of the gain and loss is nearly equal; there is no net transfer of angular momentum in the disc. We understand this balance as follows. For an $x_1$ family orbit at a given radius, the parent $x_1$ orbit is a point on the $x_1$ line. An $x_1$ orbit that is near the parent in phase-space will cross the $x_1$ line over a radial period. When the orbit has apocentres that trail the bar, it will gain angular momentum, and begin to precess forward towards the bar. Conversely, when the orbit has apocentres that lead the bar, it will lose angular momentum and precess backwards toward the bar. The positive-negative signature on either side of the $x_1$ track is not different orbits, but the same orbits in different phases of its librating precession.

However, apart from the steady-state phase, the positive-negative regions on either side of the $x_1$ track do not balance. Orbits that lose angular momentum dominate the growth phase, again located in $\rapo-\vapo$ space at higher velocities than the $x_1$ track. In the growth phase (panel b), the loss of angular momentum in the disc outpaces the gain, so the disc loses net angular momentum to the dark matter halo. During the assembly phase (panel a), the loss of angular momentum again outpaces the gain: angular momentum is still being lost to the halo. However, we also see a larger region in radius of angular momentum loss, centred at $R\seq2.4a$, which we attribute to CR, as determined determined by the monopole part of the gravitational potential. Clearly, CR is an important channel for angular momentum transfer during the assembly phase, but its contribution to bar evolution at later times becomes less important, including during the growth phase.

Panels d-i of Figure~\ref{fig:cusp_lztransfer} divide the total disc into its bar trapped and untrapped components. In the bar panels, the trapped component is confined to smaller radii at all velocities, but also does not dominate the angular momentum transfer. Qualitatively, the trapped disc behaves the same way as the total disc, losing angular momentum during the assembly and growth phases, with its angular momentum remaining roughly constant during the steady-state phase. It is apparent that the location of the commensurabilities correlates with the regions of angular momentum transfer. Comparing with the angular momentum versus time in Figure~\ref{fig:cusp_angmom} (panel c), we can see that the addition of new material dominates the increase of angular momentum. The material that is already trapped transfers angular momentum to the dark matter halo and the bar slows. During the growth phase, the trapped orbits almost uniformly lose angular momentum, which is reflected in the overall angular momentum of the bar, even as mass is still being added. During the steady-state phase, the low-velocity particles gain angular momentum and offset the angular momentum loss from high velocity particles, understood to be the precession of orbits that are not perfectly resonant (as above). This cancellation is consistent with the net zero change in angular momentum seen during the steady-state phase in Figure~\ref{fig:cusp_angmom} (panel c): no particles are being added to the bar and the angular momentum remains constant.

Panels g,h, and i of Figure~\ref{fig:cusp_lztransfer} illustrates orbits joining the bar. In the assembly phase (panel g), orbits lose angular momentum both at the end of the fledgling $x_1$ track, and also at the location of CR. During the growth phase (panel h), orbits just outside of the $x_{1b}$ track rapidly lose angular momentum, fuelling continued bar growth. In \citetalias{petersen18a}, we demonstrated that the primary channel for orbits to join the bar is by family switching. As the bar evolves, orbits located near the end of the bar pass through a 3:$n$ family co-located with the long-period $x_{1l}$ family. This channel makes up the bulk of the negative angular momentum seen near the end of the bar ($\sim 2a$) for the untrapped disc in panel h of Figure~\ref{fig:cusp_lztransfer}.

In the steady-state phase, both the bar (panel f) and untrapped disc (panel i) are angular momentum-neutral, gaining or losing angular momentum depending upon which side of the resonance the orbits exist. The same orbit will appear to be losing angular momentum when its apocentres lead the bar position angle and appear to be gaining angular momentum when its apocentres trail the bar position angle\footnote{If the bar were in true steady state and one waited long enough, the same orbit would do both.}.

In Figure~\ref{fig:core_lztransfer}, we show the transfer of angular momentum computed from the particles for the core simulation. Though the steady-state phase precedes the growth phase in this model, we choose to keep the columns in the same left-to-right order for ease of dynamical mechanism comparison with Figure~\ref{fig:cusp_lztransfer}. The assembly phase in the core simulation is characterised by huge amounts of angular momentum {\it gain} in the total disc (panel a), which we can see by examination of panels d and g, is attributable to the trapped component. This is in stark contrast to the cusp simulation. Thus, even though the assembly appears to be visually similar between the two models, the underlying torque channel appears to be quite different. The ratio of $x_1$ to other bar orbits differs between the two models. The `other' bar supporting orbits dominate the cusp model until the end of the assembly phase. The core transitions to being dominated by $x_1$ orbits at an earlier time. The increase in angular momentum for bar orbits in the core model during the assembly phase is likely related to the conversion of other bar supporting orbits into the longer $x_1$ family. The untrapped disc exhibits weak angular momentum transfer (panels g, h, and i), likely related to the low phase-space density at the position-velocity locus of CR ($\rapo=2a$, $\vapo=1.2$, cf. phase-space density in Figure~\ref{fig:core_density}).

During the steady-state phase in the core model (right column of Figure~\ref{fig:core_lztransfer}), trapped particles dominate the change in angular momentum for disc particles, though there is no net transfer (cf. panel d of Figure~\ref{fig:cusp_angmom} during the steady-state phase). As expected from our analysis of the cusp simulation, the $x_1$ track divides the orbits into those gaining and losing angular momentum based on their velocity relative to the $x_1$ track. The untrapped disc again transfers little angular momentum during the steady-state phase (panel i of Figure~\ref{fig:core_lztransfer}), as in the assembly phase (and unlike the steady-state phase of the cusp simulation, cf. panel i of Figure~\ref{fig:cusp_angmom}). Some untrapped orbits that reside in the same region of $\rapo-\vapo$ space participate in the same angular momentum transfer as the bar particles, despite not being clearly trapped. The growth phase of the core model (right column in Figure~\ref{fig:core_lztransfer}) exhibits essentially the same dynamical signatures as the growth phase of the cusp simulation (middle column in Figure~\ref{fig:cusp_lztransfer}): particles in both the trapped and untrapped bar participate in an exchange of angular momentum, with particles gaining (losing) angular momentum above (below) the $x_1$ track. The appearance of the $x_{1b}$ bifurcation again divides the bar and untrapped disc particles. As in the cusp simulation, the angular momentum transfer exhibited by the halo in $\rapo-\vapo$ space is ambiguous at best\footnote{$\rapo-\vapo$ space may not be the best representation for the more isotropic halo distribution. Presumably, the resonances are present but are washed out in this representation.}. Little-to-no angular momentum transfer occurs at the disc circular velocity at any given radius. In general, the angular momentum change for the cored halo is less everywhere compared to the cusp halo, reflective of the lower density in the halo at all radii relevant to the disc.

\subsection{Torque Applied by the Field} \label{subsec:fieldtorque}

The field quantities (e.g. density, forces, potential) at any given output time and for any given component are easily recovered with our simulation code \textsc{exp}. Above, we discussed the utility of examining the ensemble quantities and examining the quantities for individual particles. Likewise, we can perform an analysis on the field quantities, and try to identify channels through which angular momentum travels. Torque is applied by a lagged wake, which is a second-order secular response. Therefore, tracking the torque, which gives us insight into the wake in the different components, provides us with physical insight. We calculate the torque on any particle given the accumulated field as \begin{equation} \tau = \frac{d\Lz}{dt} = r\times {\bf F} = rF_\phi. \label{eq:torque} \end{equation} A BFE code such as \textsc{exp} allows us to easily track the forces throughout the simulation for any subset of particles. The basis may be partitioned into subsets for which the coefficients of the basis functions can be partially accumulated. Owing to the conditioning of the basis on the underlying initial distribution, this partial accumulation results in reconstructed densities that match the true distribution to within 3 per cent\footnote{We measure the relative difference between the sum of the partial component densities and the density computed for the entire distribution, finding $\approx$3 per cent variation in any given snapshot.}. We confine our subsets to be larger than 10 per cent of the total particles in a particular component to achieve a sufficient signal-to-noise ratio. In turn, this allows us to trace the ensembles responsible for a given angular momentum transfer feature.

Figure~\ref{fig:net_torque} shows the torque, $\tau$, evaluated instantaneously for each orbit in the cusp (upper block, panels a-f) and core (lower block, panels g-l) simulations at their respective radius and tangential velocity, computed as above. We have overlayed the relevant commensurabilities. We are primarily interested in the torque induced on the bar by the halo and on the bar by the untrapped disc. We inspected the opposite channels (torque induced in the halo by the bar and torque induced in the untrapped disc by the bar), which show approximately the opposite torques. We also inspected the remaining channels (untrapped disc on the halo, halo on the disc, halo on the halo, untrapped disc on untrapped disc and bar on the bar) and found them to be dominated by those in Figure~\ref{fig:net_torque}. This implies that direct torques between the untrapped disk and the halo play a negligible role in the evolution compared to those mediated by the bar.

In Figure~\ref{fig:net_torque}, Rows 1 and 2 (panels a-f) show the cusp model and Rows 3 and 4 show the core model (panels g-l).  For each pair of rows, the upper shows the torque on the bar from the halo, and the lower shows the torque on bar by the untrapped disc. The halo torquing the bar is the most important channel for bar evolution, followed by the disc torquing the bar. We quantify the importance of each channel by reporting the mean torque applied to particles in the lower right of each panel. The colour bar is the normalised torque $\tau$, and is the same in each panel. Negative torque means that the component applying the torque receives angular momentum from the component. Additionally, Figure~\ref{fig:net_torque} does not contain any information about the density in phase-space, and hence does not reflect the total torque applied or the angular momentum transfer observed in the simulation, but rather provides a tool with which to understand the dynamical channels.

In the lower row of each block (panels d-f and j-l), we plot the torque on the bar by the untrapped disc. This channel illustrates that while assembly is a complex process, orbits at larger velocity than the $x_1$ track receive angular momentum that moves them toward the $x_1$ track. In an analogous plot exploring the torque applied to the bar by bar particles (not shown), we see an equally complex process, which we interpret as evidence for family switching within the bar, particularly for particles not near the $x_1$ track. During the growth and assembly phases, the disc applies a positive torque on the bar. Overall, Figure~\ref{fig:cusp_angmom} demonstrates that the angular momentum of the bar decreases during the growth phase, and remains constant during the steady-state phase. Clearly, the torque applied on the bar by the disc is being offset by a different loss channel. In the case of the growth phase, the negative torque from the halo outweighs the positive torque from the disc. However, during the steady-state phase the positive torque from the disc must be balanced by some other factor, such as orbital family switching.

The true utility of Figure~\ref{fig:net_torque} is realised by comparing it with Figure~\ref{fig:cusp_lztransfer}. The applied torque helps to clarify the channel that is responsible for the observed angular momentum transfer. Some features that we noted in the angular momentum transfer figure for the cusp simulation, Figure~\ref{fig:cusp_lztransfer}, can be interpreted in a new light. There are two possible reasons for features in the angular momentum diagram that we do not see reflected in the torque diagram. The first is that the angular momentum exchange in Figure~\ref{fig:cusp_lztransfer} is just the result of first-order effects that have not fully cancelled. One can see an example of this during the steady-state phase, where despite the magnitude of the angular momentum transfer being observed in Figure~\ref{fig:cusp_lztransfer}, no responsible channel for a sustained torque can be seen in Figure~\ref{fig:net_torque} at those positions in $\rapo-\vapo$ space. The second reason is that the bar may be slowly rearranging itself without any net torque. In this paradigm, the conversion of orbits from the less eccentric $x_{1b}$ family to the more eccentric $x_{1}$ orbit family results in a change of the bar geometry and in the net angular momentum, but no torque is applied.

The halo torque during the growth phase is the dominant channel causing the bar particles to lose angular momentum, as expected from dynamical friction \citep{tremaine84}. The torque occurs across a range of velocities, but the magnitude is largest at velocities higher than the $x_1$ track, consistent with the weak angular momentum transfer in Figure~\ref{fig:cusp_lztransfer}. However, during the steady-state phase in the cusp simulation (panel c), the halo torque becomes positive, centred on the $x_1$ track, which we see reflected in the angular momentum change in the trapped component during the steady-state phase. This change to positive signals the end of bar evolution in the cusp simulation, when the halo is no longer able to exert a negative torque on the bar. The torque that the untrapped disc attempts to add to the bar has also been reduced, and is confined to smaller radii. As the main source of angular momentum is the outer disc, when the torque and angular momentum transfer in and on the disc, both by the halo and bar, goes to zero at late times, we may reasonably conclude that the outer disc can no longer efficiently contribute angular momentum to the bar.

The lower block of Figure~\ref{fig:net_torque}, panels g-l, shows the same analysis as in the upper block, but for the cored halo model. As in the $\Lz$ transfer analysis of Section~\ref{subsec:particletorque}, we see dynamical consistencies between the two simulations as well as some key differences. The formation mode for the bar differs markedly between the cusp and core simulations. We noted in Figure~\ref{fig:core_lztransfer} that the bar appeared to undergo a much more violent formation phase, as evidenced by a large gain in angular momentum by the bar particles. In the core simulation, we find that the largest source of torque on the bar comes from the untrapped disc rather than from the dark matter halo as in the cusp simulation. {\it This indicates that the formation scenario for the bars in the cusp and core simulations is in fact different. In the cusp simulation the halo mediates bar assembly, while in the core simulation the disc mediates bar assembly.} We can see this contrast in the torque diagrams for the two simulations during the growth phases. In the cusp simulation, the halo torque always occurs more interior than the outer disc torque, implying that the cusp efficiently couples with the disc and accepts angular momentum from ILR. Without the cusp, the cored halo is unable to accept angular momentum as efficiently, and must turn to the outer disc.

The importance of the torque applied by the disc persists in the other phases as well. The magnitude of the torque applied to the bar by the untrapped disc particles (bottom row of Figure~\ref{fig:net_torque}) during the steady-state and growth phases far exceeds any other ensemble during any phase save for the torque by the untrapped disc on the untrapped disc during the assembly phase, owing to the presence of strong spiral arms during assembly. As in the cusp simulation, the growth phase is again marked by an increased torque by the bar on the untrapped disc particles, again centred on the $x_{1b}$ family (indicated by the growth panel in the core simulation for the halo-torquing-bar channel of Figure~\ref{fig:net_torque}). The sign of the torque on the bar by the halo is reversed for the steady-state and growth phases of the core simulation relative to the cusp simulation (the cored halo induces negative torque on the bar during the steady-state phase and positive torque during the growth phase, while the cusped halo induces negative torque during the growth phase and positive torque during the steady-state phase). However, the magnitude of the torque in the core simulation is a factor of ten less than that in the cusp simulation, so the reversed signs may indicate just how little role the halo plays in the direct dynamics of the core simulation relative to the cusp simulation.

\subsection{Summary} \label{subsec:summary}

\begin{figure*}\centering \includegraphics[width=5.5in]{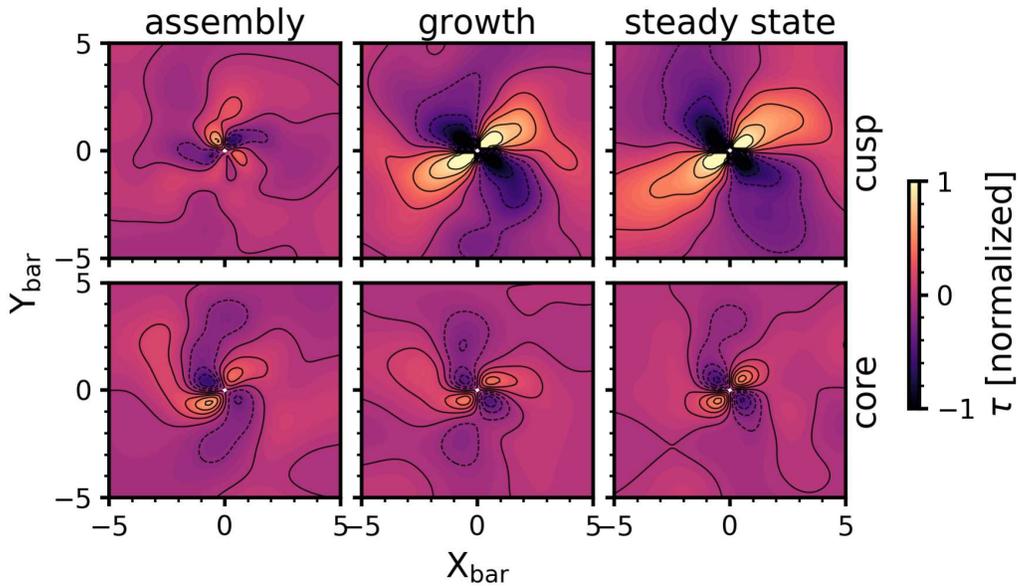} \caption{The torque field from the halo, computed from the halo basis as $rF_\phi$. The torque has been normalised such that all panels are on the same scale. The disc bar is always oriented with the $x$ axis. The columns correspond to the assembly, growth, and steady-state phases from left to right. The upper row is for the cusp simulation, and the lower row is for the core simulation. \label{fig:comp_torque}} \end{figure*}

While the previous sections offered a detailed look at two simulations, we may readily summarise the angular momentum economy using our conjecture that a halo with higher central density (the cusp model) will control the evolution, while a halo with a lower central density (the core model) will seek other channels, using the map of the torque applied by the halo wake.

Contrasting torque fields in the halo neatly summarises the differences in evolution between the two models. Figure~\ref{fig:comp_torque} shows the torque field during the three evolutionary phases (columns) for both models (rows). In contrast to the instantaneous torque in Figure~\ref{fig:net_torque}, the torque fields contain no phase-space density evolution. Their interpretive utility lies in the characteristics of the torque field. A halo that is maximally torqued will have a wake or response that resembles a perfect quadrupole with no lag (position angle variation of the quadrupole with radius, or position angle different from that of the bar). A halo wake or response that is perfectly aligned with the bar will exert no net torque. The torque field will be positive ahead of the bar and negative behind the bar with a net cancellation.

We identify the magnitude and the lag as the quantities of interest. We immediately see that the steady-state phase of the cusp simulation is both the strongest and most consistent with a pure $m=2$ disturbance, suggesting that the halo has been maximally torqued, in line with our interpretation that the cusp halo both accepted significant $\Lz$, and is not able to accept further $\Lz$. That is, rearrangement by net second-order secular torque has saturated the halo such that the phase-space density around the resonance may not accept more $\Lz$. In contrast, the growth phase in the cusp has both smaller magnitude and a lag at larger radii, consistent with ongoing angular momentum transfer; and the steady-state phase in the core has significantly smaller magnitude. The assembly phase shows a forming $m\seq2$ disturbance in both the cusp and core, where the cusp is clearly stronger even at these early times. Taken together, the torque fields support our conclusion that bar formation may be mediated by either the halo or the outer disc, depending upon the central density of the halo.

\section{Discussion} \label{sec:discussion}

We describe and summarise the key, contrasting results of our analyses of bar-driven evolution in cusp and cored halo models (Section~\ref{subsec:implications}) and outline some possible observational diagnostics (Section~\ref{subsec:observations}).

\subsection{Implications for galaxy evolution} \label{subsec:implications}

We have shown that the evolutionary phases of bars are correlated with the interplay of distinct dynamical mechanisms. The clear differences in the patterns of angular momentum transfer during the different evolutionary phases implies that secular evolution is a diverse and complex collection of interlocking couplings. In both simulations, the growth phase best corresponds to the classic picture of bar-induced secular evolution. The growth phase is the most dynamically consistent in both simulations, with a similar behaviour observed in the gross properties, the angular momentum transfer, and the torque.

The difference in the torque of the dark matter halo on the bar reflects the magnitude of the inner halo response between the cusp and cored halos. Overall the cusp and core simulations find different resonant channels to transport angular momentum. This is a direct consequence of both different potential and phase-space distributions. In the cusp simulation, the halo response to the bar is stronger, while in the core simulation, the untrapped disc response plays a larger part in the dynamics than the lower-density halo response. Further work is needed to understand the phase-space distributions that will result in the halo controlling the angular momentum transfer versus the disc in the real universe, but this difference may provide an observational diagnostic. Further, the presence of a steady-state phase in both simulations suggests that bars may be able to reach a stable configuration.

Using a large sample of galaxies from Galaxy Zoo, \citet{kruk18} found the discs of barred galaxies to be appreciably more red than unbarred galaxies. As a whole, these barred galaxies may actually be in a steady-state, rather than continuously evolving. The bars would therefore be long-lived. However, the difference in the steady-state phases between the cusp and core simulation means that even if a steady-state phase appears likely for observed bars, it may either be a final configuration (as in the cusp simulation), or a transient equilibrium that will evolve slowly until new channels can open for its growth (as in the core simulation). Currently, we cannot propose any metrics with which to separate the stable and transient equilibria.

The approximately one Gyr duration of the assembly phase means that it may be possible to catch bars in the act of secular formation. Inspection of the face-on surface density of the disc during assembly shows that a visually classifiable bar feature, e.g. apparent moderate ellipticity, is apparent from the outset of the assembly phase. As observations reach to higher redshifts many observed barred galaxies may still be in the assembly phase. Depending upon the interior density of the dark matter halo, the assembly phase may rearrange high fractions ($>50$ per cent) of the stellar disk joining the bar, taking part in transient spiral arms, or being forced outward to larger radii by the formation of the bar. If the bar assembly proceeds from an unstable disc, the power in the bar formation process may be able to modify metallicity gradients, and explain the observed range of metallicity gradients in barred galaxies. Further theoretical work is required to understand the physics of mixing in disc galaxies, including the `radial migration' mechanism \citep{sellwood02}, as well as bar-driven radial and vertical mixing \citep{petersen18e}.

Although the dynamical implications of standard bar formation predict gradients owing to formation and evolution, no strong consensus from observations about the effect of bar formation and evolution on galaxies exists. For example, \citet{sanchez14} used integral field spectroscopy from the CALIFA survey to analyse gas-phase metallicity gradients for a modest sample (N$=62$) of nearby barred and unbarred galaxies. They found no difference in metallicity or age gradients for the barred and unbarred samples. Conversely, \citet{frasermckelvie19} provided a more subtle view of bar-driven evolution using a moderately larger sample of barred galaxies (N$=128$) obtained as part of the SDSS-IV MaNGA Galaxy Survey \citep{bundy15,drory15}. They find that within the barred region, the age and metallicity gradients are flatter than in disc-dominated regions misaligned with the bar at radii within the bar. This result suggests that the azimuthal-averaging of barred galaxy gradients could play a role in removing the signal of bar-induced radial mixing: by azimuthally averaging disc-dominated regions with large metallicity gradients into bar-dominated regions with flat metallicity gradients, one may miss the difference between barred and unbarred samples (particularly in the case of weak bars). The size of the observational samples and apparently contradictory results suggests that one needs further observational work to overcome potential sources of observational bias, preferably with integral field spectroscopy of barred and unbarred galaxies extending to several bar radii.

\subsection{Utility for Observations} \label{subsec:observations}

The distribution of the disc in $\rapo-\vapo$ position-velocity space is a useful tool for learning about the evolutionary state of observed barred galaxies. Out of all our diagnostics, the density plots Figures~\ref{fig:cusp_density} and \ref{fig:core_density} may be directly constrained for observed galaxies. Unfortunately, these are arguably the least informative of all our diagnostics for galaxy evolution, and are best interpreted with a wealth of other information that only simulations can provide. In spite of this, we propose that the density plots in radius and velocity space can be used to rule out different scenarios as follows.

In both the cusp and core models, we observe three distinct phases: assembly, growth, and a steady-state. The relevant question then is {\it during what phase of evolution do we observe real bars?} All three phases have extended durations during our simulation, $\Delta T_{\rm phase}\approx2$ Gyr for a MW-like galaxy. During transitions between the phases a blended version of the trends seen during the clear phases can lead to ambiguous signatures. However, if a local minima or `gap' is observed between the high velocity and low velocity peaks of the density, the bar is likely to be dynamically evolved.

With IFUs one can construct the density $\rapo-\vapo$ plane to look for the influence of angular momentum transfer in real observed galaxies. The effects are subtle and rely on both high spatial resolution ($\sim1$kpc) and high velocity resolution ($\sim10\kms$). However, instruments such as MUSE on the VLT may have the ability to create a useful $\rapo-\vapo$ diagram for some nearby barred galaxies and compare them with the density plots presented in this paper to look for (1) radius-velocity separation of the bar and untrapped disk, and (2) the pattern speed of the bar, which is the point where the bar feature reaches its largest radial extent, and (3) any suggestion of gaps owing to commensurabilities in the outer disc, which may determine the location of CR or other strong resonances.

\section{Conclusion} \label{sec:conclusion}

We analyse secular evolution in barred galaxy simulations based on the gross properties of distinct subsets of orbits defining the bar, the untrapped disc, and the dark matter halo.  We independently isolate components that dominate the instantaneous torque through time. We associate bar evolutionary phases with changes in the observed distribution of torques in $\rapo-\vapo$ space. The framework is generally applicable to simulated model galaxies. We presented two models roughly consistent with the MW.

The main results of this paper are as follows: 
\begin{enumerate} \renewcommand{\theenumi}{(\arabic{enumi})} 
\item Careful accounting of the angular momentum budget reveals that the angular momentum economy of the bar-disc-dark matter halo system is composed of several key transport channels: the assembly of the bar, the friction imposed by the halo and disc on the bar, and the coupling of the disc to the halo. Resonances control the transfer of angular momentum from the bar pattern to the halo, as well as from the disc to the halo directly. 
\item The assembly of the bar is marked by individual orbits losing angular momentum to join the bar pattern after being torqued by the dark matter halo or the untrapped disc. 
\item A long-lived bar will reach net zero angular momentum transfer where the orbits leading and trailing the bar exert equal but opposite torques, reaching a steady-state in the cusp simulation where the bar pattern no longer slows (Figures~\ref{fig:cusp_lztransfer} and \ref{fig:core_lztransfer}). 
\item The coupling of the bar to the disc and the dark matter halo is different in the cusp and core simulations, with the cusp simulation using the halo as the primary driver of evolution, and the core simulation using the disc as its primary driver (Figures~\ref{fig:mean_lz}, \ref{fig:net_torque}, and \ref{fig:comp_torque}).
 \end{enumerate}

The coupling between the disc and the dark matter halo requires a transport `channel', which commensurate orbit families provide. Commensurabilities (resonances) provide secular channels to funnel angular momentum from sources to sinks. Secular evolution rearranges both energy and angular momentum, slowly changing the gravitational potential.  As the potential changes, a non-commensurate disc orbit becomes commensurate, promoting further evolution. Once a disc orbit becomes a commensurate orbit in the frame of the bar, it will efficiently couple with the halo and donate angular momentum to the halo.

In short, the sources of angular momentum seek to find a sink in which to donate their angular momentum. The dominant sinks are halo-model dependent. The bar is both an efficient sink (at early times during assembly) and source (during secular growth). The bar may reach a limit in its ability to catalyse angular momentum transfer from the disc to the halo (the steady-state phase), which may be transient or long-lived. We identify the transfer of angular momentum from the disc to the bar at early times, which torques up the bar. The bar then transfers the angular momentum to the halo, and then the model arrives at an equilibrium. The outer disc plays little role in the long-term evolution of the system. In particular, we find little role beyond CR for the disc during the growth and steady-state phases, which we argue are the most applicable phases to galaxy evolution in the real universe. It is likely that bars and galactic discs spend a larger fraction of their lifetimes in the growth or steady-state phases.

The technique presented in \citetalias{petersen18a} is instrumental in identifying commensurate orbits. To analyse the potential that gives rise to the commensurate orbits, we require the harmonic analysis of \citetalias{petersen18c}. All three techniques provide different but complementary methods for understanding evolution in simulations of barred galaxies, as well as its observational applications. Using the information provided in \citetalias{petersen18a}, we associate different features in the torque $\tau$ exerted by wakes and the change in orbital angular momentum $\Delta\Lz$ with different commensurate orbits that will not be readily identified through frequency analysis. In this work, the identification of orbits associated with the bar allows for tracking $M_{\rm family}$, $L_{z, {\rm family}}$, and $R_{99, {\rm family}}$. Additionally, the ability to compute $\Delta \Lz$ and $\tau$ for subsets of particles removes much of the anecdotal and circumstantial evidence for the channels of angular momentum flow in barred disc galaxies. We place clear constraints on angular momentum signatures during different phases of secular evolution.

The instantaneous torque applied by different subsets of particles is a powerful dynamical tool that can explain the observed $\Lz$ changes of individual and localised ensembles of orbits. This work complements other techniques proposed to analyse the evolution of barred systems, namely orbital decomposition \citepalias{petersen18a} and harmonic analysis \citepalias{petersen18c}. These techniques jointly explain the microphysics of the system, like the behaviour of individual orbits, and the macrophysics, like the gross properties of the observable collective features.

\section*{Acknowledgements} We thank the anonymous referee who strengthened the results of this paper.

\bibliography{PetersenMS}

\end{document}